\documentclass[12pt,preprint,natbib]{emulateapj}

\newcommand{\be} {\begin{equation}}

\def\uu {4U\,0142$+$614}
\def\ee {1E\,1048.1$-$5937}
\def\kes {1E\,1841$-$045}
\def\aa {1E\,1547$-$5408}

\def\rxs {1RXS\,J1708$-$4009}
\def\xte{XTE\,J1810--197\,}

\def\wes{CXOU\,J1647--4552}
\def\ea {1E\,2259$+$586}
\def\sgrb{SGR\,1900+14}
\def\sgra{SGR\,1806--20}

\def\sgrd{SGR\,1627--41\,}
\def\sgre{SGR\,0501+4516}
\def\sgrf{SGR\,0418+5729}

\newcommand{\XMM}{{\em XMM--Newton}}

\newcommand{\RXTE}{{\em R}XTE}
\newcommand{\swift}{{\em Swift}}
\newcommand{\fermi}{{\em Fermi}}
\newcommand{\INT}{{\em INTEGRAL}}
\newcommand{\bc}{\begin{center}}
\newcommand{\ec}{\end{center}}
\def\ltsima{$\; \buildrel < \over \sim \;$}
\def\lsim{\lower.5ex\hbox{\ltsima}}
\def\loe{\lower.5ex\hbox{\ltsima}}
\def\gtsima{$\; \buildrel > \over \sim \;$}
\def\gsim{\lower.5ex\hbox{\gtsima}}
\def\goe{\lower.5ex\hbox{\gtsima}}

\def\ltsima{$\; \buildrel < \over \sim \;$}
\def\lsim{\lower.5ex\hbox{\ltsima}}
\def\loe{\lower.5ex\hbox{\ltsima}}
\def\gtsima{$\; \buildrel > \over \sim \;$}
\def\gsim{\lower.5ex\hbox{\gtsima}}
\def\goe{\lower.5ex\hbox{\gtsima}}

\def\ergs {erg\,s$^{-1}$}
\def\ergscm2 {erg\,s$^{-1}$cm$^{-2}$}
\def\ss {s\,s$^{-1}$}
\def\cm2 {cm$^{-2}$}
\def\arcsec{$^{\prime\prime}$}

\shortauthors{Fermi-LAT collaboration}
\shorttitle{Fermi-LAT observations of magnetars}

\begin{document}
\title{SEARCH FOR GAMMA-RAY EMISSION FROM MAGNETARS WITH THE {\em FERMI} LARGE AREA TELESCOPE}

\author{
A.~A.~Abdo\altaffilmark{1,2}, 
M.~Ackermann\altaffilmark{3}, 
M.~Ajello\altaffilmark{3}, 
A.~Allafort\altaffilmark{3}, 
L.~Baldini\altaffilmark{4}, 
J.~Ballet\altaffilmark{5}, 
G.~Barbiellini\altaffilmark{6,7}, 
M.~G.~Baring\altaffilmark{8}, 
D.~Bastieri\altaffilmark{9,10}, 
R.~Bellazzini\altaffilmark{4}
% ...
}
\altaffiltext{1}{Space Science Division, Naval Research Laboratory, Washington, DC 20375, USA}
\altaffiltext{2}{National Research Council Research Associate, National Academy of Sciences, Washington, DC 20001, USA}
\altaffiltext{3}{W. W. Hansen Experimental Physics Laboratory, Kavli Institute for Particle Astrophysics and Cosmology, Department of Physics and SLAC National Accelerator Laboratory, Stanford University, Stanford, CA 94305, USA}
\altaffiltext{4}{Istituto Nazionale di Fisica Nucleare, Sezione di Pisa, I-56127 Pisa, Italy}
\altaffiltext{5}{Laboratoire AIM, CEA-IRFU/CNRS/Universit\'e Paris Diderot, Service d'Astrophysique, CEA Saclay, 91191 Gif sur Yvette, France}
\altaffiltext{6}{Istituto Nazionale di Fisica Nucleare, Sezione di Trieste, I-34127 Trieste, Italy}
\altaffiltext{7}{Dipartimento di Fisica, Universit\`a di Trieste, I-34127 Trieste, Italy}
\altaffiltext{8}{Rice University, Department of Physics and Astronomy, MS-108, P. O. Box 1892, Houston, TX 77251, USA}
\altaffiltext{9}{Istituto Nazionale di Fisica Nucleare, Sezione di Padova, I-35131 Padova, Italy}
\altaffiltext{10}{Dipartimento di Fisica ``G. Galilei", Universit\`a di Padova, I-35131 Padova, Italy}
% ...

\begin{abstract}

We report on the search for 0.1--10\,GeV emission from magnetars in 17
months of \fermi\, Large Area Telescope (LAT) observations.  No
significant evidence for gamma-ray emission from any of the
currently-known magnetars is found.  The most stringent upper limits
to date on their persistent emission in the \fermi\ energy range are
estimated between $\sim10^{-12}-10^{-10}$\ergscm2 , depending on the
source. We also searched for gamma-ray pulsations and possible
outbursts, also with no significant detection. The upper limits
derived support the presence of a cut-off at an energy below a few MeV
in the persistent emission of magnetars.  They also show the likely
need for a revision of current models of outer gap emission from
strongly magnetized pulsars, which, in some realizations, predict
detectable GeV emission from magnetars at flux levels exceeding the
upper limits identified here using the \fermi-LAT observations.
  
\end{abstract}


\section{INTRODUCTION}
\label{intro}

Magnetars are isolated neutron stars whose emission is thought to be
powered by their magnetic energy.  They are discovered either through
their bursting activity (and in this case named Soft Gamma Repeaters;
SGRs; Kouveliotou et al. 1998) or by their strong persistent soft X-ray
emission (then named Anomalous X-ray Pulsars; AXPs; Mereghetti \&
Stella 1995). In recent years SGRs and AXPs have been
recognized as part of the magnetar class, with the discovery of
many AXPs and SGRs showing common characteristics and properties (e.g. Kaspi et al. 2003; Rea et
al. 2009; Mereghetti et al. 2009; Israel et al. 2010).

Their X-ray luminosities are typically $10^{33} - 10^{36}$\ergs . They
have spin periods between 2--12\,s and period derivatives in the
$10^{-13} - 10^{-11}$\ss range.  In most of the cases, the magnetic fields
derived from their spin periods and period derivatives, assuming only
magnetic dipolar losses as is usually done for normal pulsars, are
inferred to be $\sim 10^{14}-10^{15}$\,Gauss. These high B fields, in particular their toroidal components, 
impose tremendous stresses on neutron star crusts, and thereby are
believed to be the ultimate energy source of magnetar emission (Duncan
\& Thompson 1992; Thompson \& Duncan 1993). The strong soft X-ray
emission can be empirically modeled with a black body ({\it
  kT}$\sim$0.3-0.7\,keV), plus a power law ($\Gamma\sim$1.5-4; see
e.g. Mereghetti 2008 for a review), although recently more physically
based models have been developed to account for their X-ray spectra (e.g. Lyutikov \& Gavrill 2006; Fernandez \& Thompson
2007; Rea et al. 2008; Zane et al. 2009).  In the past years,
magnetars have been discovered as persistent hard non-thermal X-ray
sources, emitting up to $\sim$250\,keV (e.g. Kuiper et al. 2004, 2006;
G\"otz et al. 2006; den Hartog et al. 2008; Rea et al. 2009; Enoto et
al. 2010).

Our current knowledge of their spectra at much higher energies
($>$0.5\,MeV) is very limited. Archival studies of {\em COMPTEL}
observations were used to place upper limits on the emission of a few
magnetars in the 0.75-30\,MeV range, of $\sim10^{-10}$\ergscm2 at a
2$\sigma$ level (Kuiper et al. 2006).  Very poor so far is the knowledge 
concerning their behavior at energies $>$30\,MeV (Heyl \& Hernquist 2005), a band of interest
given model predictions of measurable synchrotron/curvature emission (Chang
\& Zheng 2001; Zhang \& Cheng 2002). 

The Large Area Telescope (LAT), the main instrument on board the
{\em Fermi} Gamma-ray Space Telescope, launched on 2008 June 11th, is
the most sensitive telescope to date in the GeV energy range. 
We present in this Letter results of a search for emission in the GeV domain 
from the first 17 months of \fermi-LAT observations of magnetars\footnote[1]{We note that during the submission phase of this work another paper has been published reporting upper limits on one of the sources reported in this paper (Sasmaz Mus \& Gogus 2010).}.


\begin{center}
\begin{deluxetable*}{lccccclllc}
\tabletypesize{\scriptsize}
\tablecaption{{\em Fermi}-LAT upper limits on magnetars obtained from likelihood analysis.}
\tablewidth{0pt}
\tablehead{
\colhead{Source} &  \colhead{d$^{*}$} & \colhead{log(B)} &  \colhead{log(L$_{\rm X}$)$^{*}$} &  \colhead{log(L$_{\rm rot}$)} & \colhead{TS} & \colhead{0.1--10\,GeV} & \colhead{0.1--1\,GeV} & \colhead{1--10\,GeV} & \colhead{1FGL srcs}  \\
  & \colhead{kpc} & \colhead{Gauss} & \colhead{erg~s$^{-1}$} & \colhead{erg~s$^{-1}$} &  & ($\Gamma=2.5$) & ($\Gamma=1.5$) & ($\Gamma=3.5$) & within 3$^{\circ}$
}
 \startdata
  \ee &  3.0 & 14.78 & 34.00 & 33.90 & 0.0 &   $<$5.3 (12.0)  &  $<$3.9 (7.7)  &  $<$1.7 (0.7)   & 7 \\
 \sgrb &  15 & 14.81  & 35.44 & 34.34 & 0.0 &  $<$0.4 (0.9)  &  $<$0.8 (2.0) &  $<$0.6 (0.2)  & 5 \\
 \sgrf & 2.0 & $<$12.70 & 31.77 & $<$29.47 & 2.3 & $<$0.4 (0.9)&  $<$0.2 (0.4) &  $<$0.1 (0.04)    & 2 \\
 \sgra & 8.7 & 15.15 & 35.21 & 34.40 & 2.8 &  $<$0.6 (1.4) &  $<$0.5 (0.9) &  $<$0.12 (0.05)     & 1\\
  \uu & 5.0 & 14.11 & 35.32 & 32.10 & 3.6 &  $<$0.9 (2.0)  &  $<$0.5	 (0.9) &  $<$0.3 (0.11) & 1\\
  \kes & 8.5 & 14.85 & 35.34 & 32.99 & 7.5 &  $<$3.0	 (6.0)  &  $<$6.3 (13.0)  &  $<$2.4 (0.92)    & 8 \\
 \xte & 4.0 & 14.46 & 33.58 & 33.60 & 13.1 &  $<$5.0 (10.0)  &  $<$12.0 (23.0)  &  $<$2.0 (0.7)    & 7\\
  \ea & 3.0 & 13.76 & 34.43 & 31.70 & 15.6 &  $<$1.7 (3.9)  &  $<$0.6 (1.0)  &  $<$0.63 (0.24)   & 2 \\
\sgre & 5.0 & 14.23 & 34.77 & 33.49 & 16.3 & $<$1.9 (4.3)  &  $<$0.6 (1.0) &  $<$0.5 (0.18 )    & 1 \\
\rxs & 8.0 & 14.67 & 35.27 &  32.75 & 32.1 & $<$10.0 (20.0)  &  $<$5.0 (9.0)  &  $<$9.0 (4.0)    & 8 \\
\wes & 5.0 & 14.20 & 34.41 & 31.89 & 33.7 & 	$<$10.0 (20.0)  &  $<$10.0 (20.0)  &  $<$19.0 (7.2)    & 7\\
\sgrd &  11 & 14.34 & 33.39 & 34.63 & 36.0 & 	$<$20.0 (50.0)  &  $<$20.0 (30.0)  &  $<$5.0 (2.0)    & 8 \\
\aa & 9.0 & 14.32 & 34.16 & 35.00 & 36.2 &  $<$10.0 (20.0)  &  $<$7.9 (16.0)  &  $<$2.1 (0.8)     & 6 \\

\enddata

\label{table}
\tablecomments{Properties of the magnetars studied in this work ordered by the measured TS values derived from the binned analysis (for further info on the first 4 columns see Mereghetti (2008) and reference therein; Rea et al. (2009, 2010) for the newly discovered \sgre\, and \sgrf, respectively). The GeV upper limits are
  reported at 95\% confidence level (see Sect.\ \ref{UpperLimit} for
  details). Fluxes are in units of $10^{-11}$\ergscm2 (or
  $10^{-8}$\,photons~cm$^{-2}$~s$^{-1}$ for numbers in brackets). The
  last 4 sources and \kes\, are discussed in detail in the
  text. $^{*}$ Note that most of the sources have very variable X-ray
  luminosities, and very uncertain distances, hence those values
  should be taken as indicative.}
\end{deluxetable*} 
\end{center}

\section{OBSERVATION AND DATA REDUCTION}
\label{obs}

The data analyzed here were taken in survey mode with the \fermi\, Large Area Telescope, from 4 August 2008  until 1 January 2010 .  The \fermi-LAT telescope is
sensitive to photons with energies from about 20\,MeV to more than
300\,GeV and uses the pair conversion technique. The direction of an incident photon is derived by
tracking the electron-positron pair in a high-resolution converter
tracker, and the energy of the pair is measured with a CsI(Tl) crystal
calorimeter. The \fermi-LAT has an on-axis effective area of 8000
cm$^2$, a 2.4\,sr field of view, and an angular resolution of
$\sim$0.6$^{\circ}$ at 1\,GeV (for events converting in the front
section of the tracker).  Furthermore, an anti-coincidence detector
identifies the background of charged particles (Atwood et al.  2009).

We analyzed the data using the Fermi Science Tools {\tt
v9r15} package.  Events from the ``Pass 6 Diffuse'' event class are
selected, i.e. the event class with the greatest purity of gamma rays, having the most
stringent background rejection (Atwood et al. 2009). The
``Pass 6 v3 Diffuse" instrument response functions (IRFs) are applied
in the analysis. 
For each analyzed source we select events with energy
E$>$100\,MeV in a circular region of interest (ROI) of 10$^{\circ}$
radius. The good time intervals are defined such that the ROI
does not go below the gamma-ray-bright Earth limb (defined at 105$^{\circ}$ from the
Zenith angle), and that the source is always inside the LAT field of view, namely in a cone angle of
66$^{\circ}$.

\begin{center}
\begin{figure*}
\includegraphics[height=9.4cm]{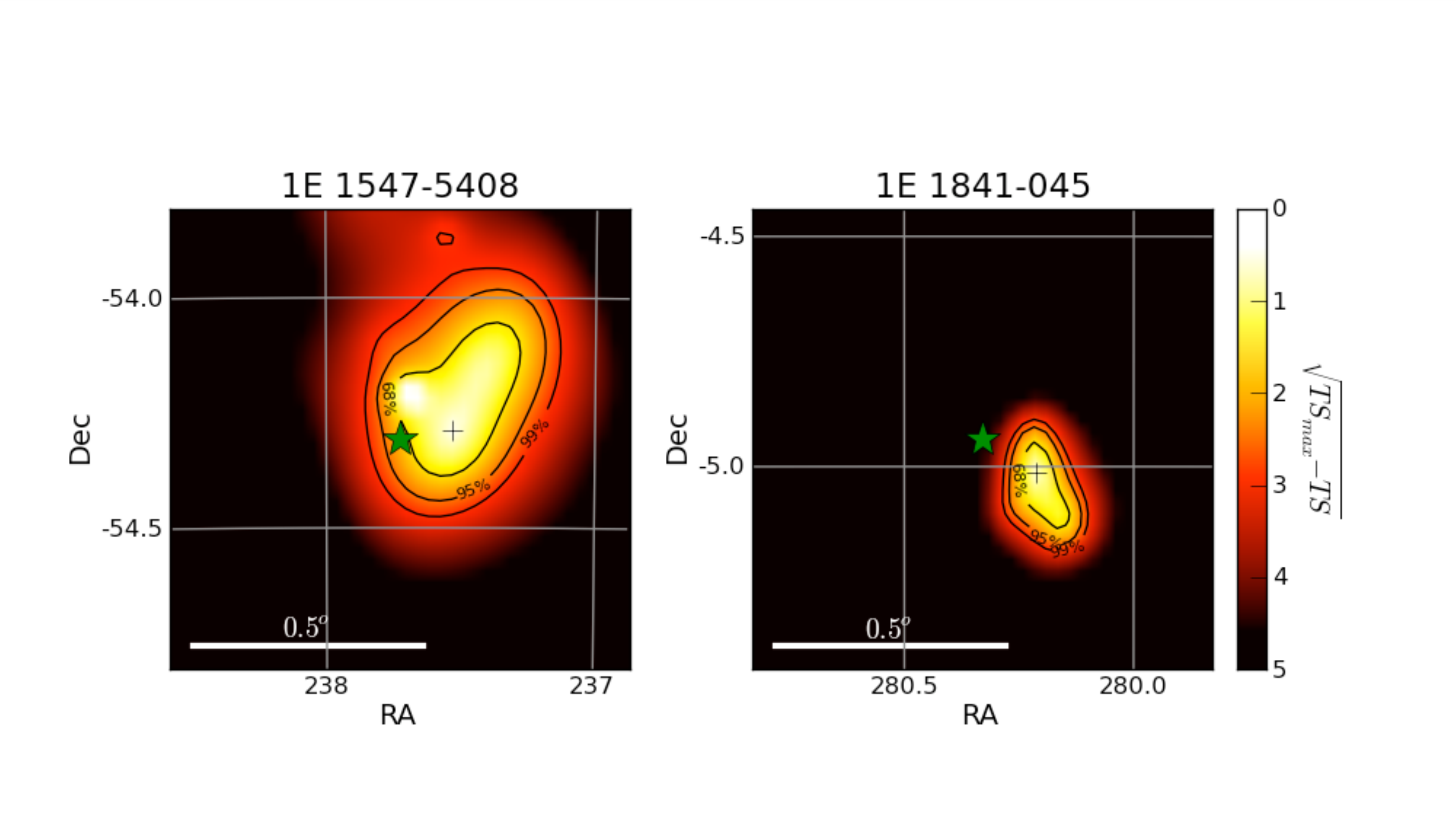}
\caption{Test Statistic maps of the \fermi-LAT fields of \aa\, and \kes\ (RA and Dec are referenced at J2000). The green stars represent the X-ray position of each magnetar. TS$_{\rm max}$ is the maximum TS value inferred around the two magnetars, measured in the position labelled by the crosses. Solid lines are the positional confidence levels around  the maximum TS value in each field of view. See text for details.}
\end{figure*}
\end{center}

\section{Likelihood analysis and results}
\label{results}

Gamma-ray emission was analyzed at the positions of 
all the magnetars known to date, excluding
yet unconfirmed candidates. Extragalactic magnetars located in the Large
and Small Magellanic Clouds are also excluded due to their large
distances and the difficulty of resolving them from their host galaxies (see Abdo et al. 2010, 2010a).  See Table 1 for the 13 selected magnetars.

The Test Statistic was employed to evaluate the significance of the gamma-ray fluxes coming from the magnetars.  The TS  value is used to assess the goodness of a fit, and it is defined as twice the difference between the log-likelihood function maximized by adjusting all parameters of the model, with and without the source, and under the assumption of a precise knowledge of the Galactic and extragalactic diffuse emission. A TS=25 roughly corresponds to a 4.6$\sigma$ detection significance (Abdo et al. 2010b).  

Binned and unbinned likelihood analysis are applied on the data, using the official tool ({\tt gtlike}) released by the \fermi-LAT collaboration. The binned likelihood uses events selected in a square inscribed inside the circular ROI (see \S\,2), aligned with celestial coordinates.

For each magnetar, a spectral-spatial model containing diffuse and
point-like sources is created, and the parameters are obtained from a maximum likelihood fit to the data.  For the Galactic
diffuse emission we use the spectral-spatial model
``gll\_iem\_v02.fit", used by the \fermi\, collaboration to build the
First \fermi\, Source Catalog (Abdo et al. 2010b; 1FGL hereafter). The
extragalactic diffuse emission was modeled as an isotropic emission using  the spectrum
described in the ``isotropic\_iem\_v02.txt" file\footnote[2]{All the data, software, and diffuse models used for this analysis are available from the Fermi Science Support Center. http://fermi.gsfc.nasa.gov/ssc/}. 
This spectrum also takes into account the residual background of charged particles in the LAT.

In the spectral-spatial model of each magnetar we fixed its position at the localization determined by X-ray observations (in all cases with uncertainties $<$2\arcsec; see Mereghetti 2008 and the McGill catalog\footnote[3]{www.physics.mcgill.ca/$\sim$pulsar/magnetar/main.html}), and also included all the point-like sources from the 1FGL list 
closer than 15$^{\circ}$. Each of those point sources was  modeled with a simple
power-law, with the exceptions of the pulsars closer than 3$^{\circ}$
from the magnetars, for which a power-law with an exponential cut-off
was used. The spectral parameters of those sources were fixed at
the 1FGL values or those from the \fermi-LAT First Pulsar Catalog (Abdo et al. 2010c), 
while the flux parameters of all the point-like sources closer than 3$^{\circ}$ to the magnetar were left free
in the likelihood fit (see also Table\,1, last column).

We modeled the magnetar emission 
using power-law  spectral distributions with two free parameters:
the flux and spectral index. 
The likelihood ratio test indicated values of TS less than 25
 for most of the analyzed magnetars (see Table\, 1). For \rxs ,
\wes , and \aa\, the calculated TS values were in the range 25--50,
while \sgrd\, and \kes\, had TS$>$70.  The latter cases are addressed in Sect.\ \ref{highTS}.

For those magnetars for which X-ray outbursts were detected during the \fermi-LAT observing period (namely \sgre, \sgrf \ and \aa; e.g. Rea et al. 2009; Esposito et al. 2010; Israel et al. 2010), we re-ran the analysis considering subsets of data taken one day, one week or two weeks around the peaks of their X-ray outbursts. All TS values during those outbursts were $<$25.

\section{Sources with high TS values}
\label{highTS}

Given the relatively high TS values found in five cases by the likelihood analysis, 
we checked whether the X-ray positions of these magnetars are compatible with the most probable origin of the gamma-ray excesses. For this purpose, we performed a localization process similar to the one used for the 1FGL catalog, 
using the {\tt pointlike} tool, which returns the TS map around each source, where the TS is calculated at any putative source position.
(see Fig.\,1 for two examples of these maps, around \aa\, and \kes ). This
tool is applied leaving as free the spectral parameters of the modeled sources within 1$^{\circ}$ of each magnetar. The results for the magnetar positions with TS$ > 25$ are summarized below.
We remember here that all these high-TS sources are in the inner Galaxy close to the Galactic plane, where the diffuse emission is strong and highly structured, and this could affect our results.

\subsection{\kes}
The {\tt gtlike} analysis of \kes\, resulted in a high TS value ($>70$). In this case the {\tt pointlike} TS map showed a new source very close to the magnetar at an angular distance of 0.11$^{\circ}$ (RA=280.23$^{\circ}$, Dec=-4.99$^{\circ}$. See Fig.\,1 right panel).
This source is not present in the 1FGL catalog, 
probably due to the longer time interval analyzed here (17 months vs. 11 months for 1FGL).
The TS value of \kes\, falls
below 25 when the new source is added to the spectral-spatial model
used for the likelihood analysis, and thus we find no evidence to claim the magnetar as a gamma-ray emitter.

\subsection{\sgrd}
The {\tt pointlike} analysis for \sgrd\, indicated that the position of the magnetar was not a maximum of TS when the coordinates of the modeled source were optimized in the fit. In particular, we found that the high TS derived by the {\tt gtlike} analysis could have been caused by the presence of the rather strong unidentified source (1FGL J1636.4-47371), which lies as close as 0.12$^{\circ}$ from the magnetar (although positionally incompatible
with it). If the spectral parameters of the modeled 1FGL sources are held fixed at their values in the 1FGL catalog, \sgrd ends up having a TS$\sim$36. This is what is reported in Table\,1.
While this is still greater than 25,  the flatness of the TS map around this source suggests that in this region the diffuse Galactic emission could be underestimated by the model adopted in the likelihood analysis.

\subsection{\rxs\ and \wes}
The {\tt gtlike} analyses of these two magnetars resulted in TS values of $\sim$30 for both sources. For each source we performed a {\tt pointlike} analysis which in both cases indicated that the position of the two magnetars were not a maximum of TS when the coordinates of the modeled source were optimized in the fit. We cannot exclude that the likelihood excesses of
 \rxs\ and \wes\  are caused by the uncertainties of the Galactic diffuse model. \\

\subsection{\aa}
\aa\ is the only source for which the TS map calculated by {\tt pointlike} indicated that the position of the magnetar was indeed consistent with a local maximum of TS (see Fig.\,1 left panel).  In particular, \aa\, has a TS$\sim$35, and it is observed inside the 95\% positional error contour around the TS$_{\rm max}$ of the field. 
With the current \fermi\, observations a firm association between this excess and \aa\, can not be made. 
Furthermore, we found that the TS of \aa\ falls below 20 if the level of the Galactic diffuse emission is increased by only 2\%

\section{Upper limits evaluation}
\label{UpperLimit}

Before starting with the upper limit determination, we note that for all but one magnetar, the local maximum of TS was not coincident with the magnetar position. Furthermore, by increasing the level of the Galactic diffuse emission by 2-4\%, all of the TS values determined in \S\,3 would decrease below 20. 
These percentages are well inside the systematics of the assumed Galactic diffuse emission model (see the cases of the supernova remnants W51C and W49, and  \S\,4.7 of the 1FGL catalog; Abdo et al. 2009, 2010d, 2010a).

The discovery of GeV gamma-rays from magnetars would  have major implications, hence would require very strong evidence. The evidence so far does not seem to reach more than the circumstantial level, and while the \fermi-LAT exposure continues to accumulate on these sources, we find it appropriate for the time being to report only upper limits.

The upper limits are evaluated by applying the binned likelihood analysis, and using the spectral-spatial models described above. We derived 95\% flux upper limits by fitting a
point source at the X-ray magnetar position, for which we
increase the flux until the maximum likelihood decreases by 2.71/2 in logarithm.

In the 0.1--10\,GeV energy range we fix the photon index value of
the magnetars to 2.5, which is the mean of the
photon indexes obtained by the previous likelihood analyses.  The other
two upper limits are evaluated using spectral index values that mimic a
cutoff in the spectrum at $\sim1$\,GeV, as common in pulsar spectra. Accordingly, in the range 0.1--1\,GeV we
fix the spectral index to 1.5, while for 1--10\,GeV it is
set as 3.5.

The uncertainties of the \fermi-LAT effective area and of the Galactic diffuse emission are the two main sources of systematics that can affect the evaluation of the upper limits. We estimated the effect of these systematics by repeating the upper limits analysis using modified instrument response functions that bracket the ``Pass 6 v3 Diffuse"  effective areas, and changing the normalization of the Galactic diffuse model artificially by $\pm 6\%$. The results of this analysis are reported in Table 1.

\section{Timing analysis}

A timing analysis was performed for each of the 13 magnetars studied in this work. With this aim we used the X-ray data available for these objects to build their ephemerides to fold the \fermi-LAT data, or we searched around their X-ray periods when a long-baseline ephemeris could not be derived.  In particular, using \RXTE\, and \swift-XRT data, ephemerides\footnote{Only in a few cases a single ephemeris could be derived over the entire time-baseline, while in other cases 4-5 different ephemerides were needed to cover the whole \fermi-LAT data span.} have been derived for \uu, \ea, \ee, 1RXS\,J1708 $-$4009, \kes, \aa, and \sgre\,  (Dib et al. in prep; Israel et al. 2010; Bernardini et al. in prep; Rea et al. 2009; Rea et al. in prep.).  
For each of the other magnetars, an ephemeris valid throughout the 17 months of \fermi-LAT observations is not derivable either given the paucity of X-ray observations or because the source is too dim to have long-term measurements of its spin period. 
For these we searched directly in the gamma-ray data, performing a semi-blind search around plausible values of spin period and its derivative (see Mereghetti 2008). With the help of PRESTO software (Ransom 2001), we also tried to improve the signal including trials for the second derivative of the period.
No significant signal has been detected either searching in \fermi-LAT data around the X-ray period, or, when possible, folding at the X-ray ephemeris derived from current
X-ray monitoring observations.

\begin{center}
\begin{figure}
\includegraphics[width=9cm]{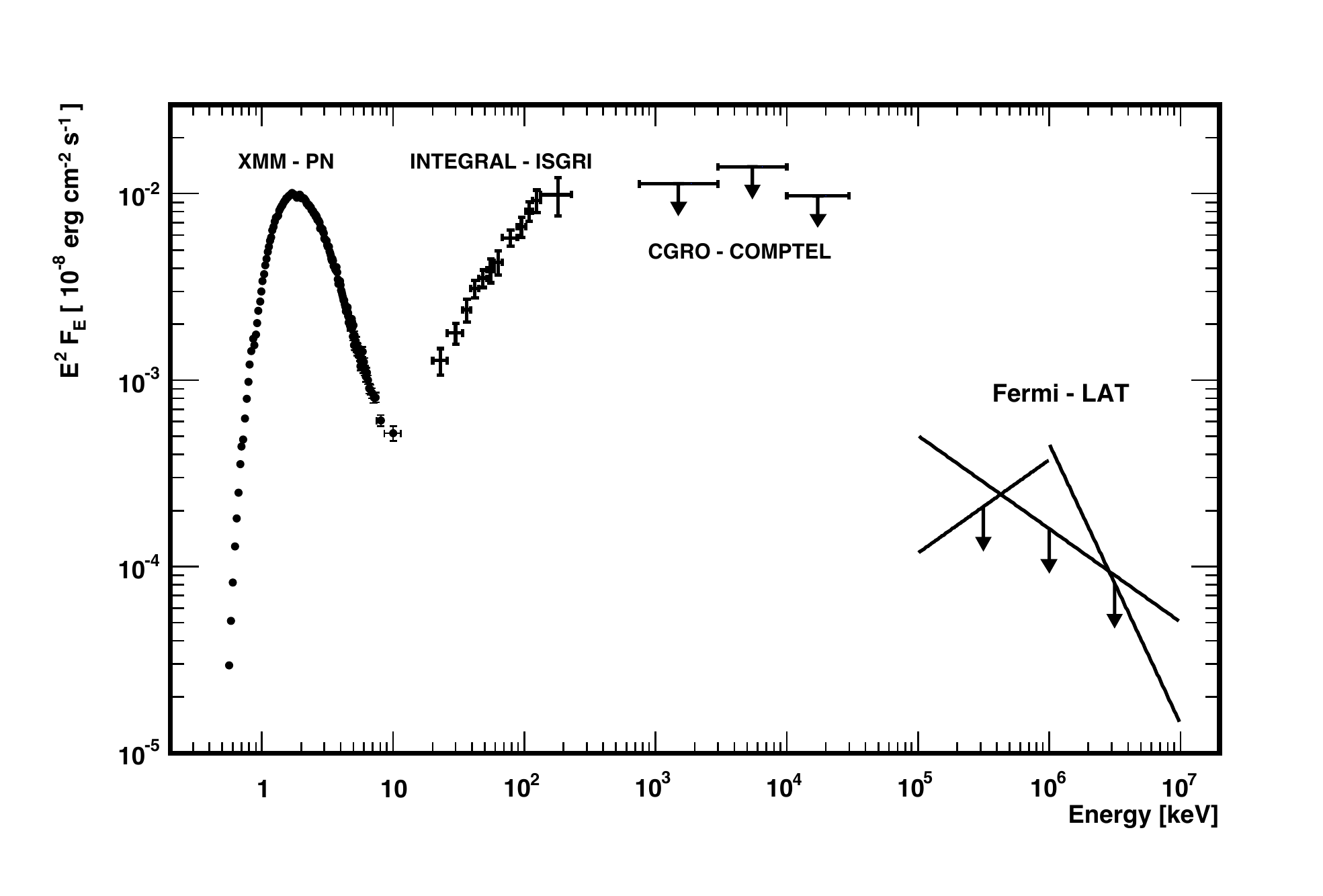}
\caption{Multi-band E$^2$F$_E$ spectrum of \uu. The 0.1--200 keV data are from  \XMM-PN and \INT-ISGRI (Rea et al. 2007a; den Hartog et al. 2008; Gonzalez et al. 2010), on which we over-plot the 2$\sigma$ {\em COMPTEL} upper limits (den Hartog et al. 2006; Kuiper et al. 2006) and the  \fermi-LAT limit (this work) with the assumed power-law spectrum with photon index values of 2.5, 1.5 and 3.5 in  the 0.1--10\,GeV, 0.1--1\,GeV and 1--10\,GeV bands, respectively (see also Table 1).}
\end{figure}
\end{center}

\section{DISCUSSION}
\label{discussion}

In this paper we searched for GeV emission from magnetars using the most sensitive data to date. We did not find evidence beyond reasonable doubt that would allow us to claim the detection of any of these magnetars.
In a few cases, putative detections were marked for further studies, but they are not significant enough to claim a new population of gamma-ray emitters.  

For all of the studied magnetars we calculated the deepest upper limits derived to date in the 0.1--10\,GeV energy range. In Fig.\,2 we show the 0.1\,keV--10\,GeV multi-band spectrum of \uu, the
persistent magnetar having the brightest emission and steepest spectral decomposition in the hard X-ray band.  Comparing the
\fermi-LAT upper limits to the hard X-ray measured fluxes for all the studied magnetars, it is clear that the spectral energy distribution of these objects should necessarily have a cut-off below the MeV band, as already pointed out for a few sources by {\it COMPTEL} observations (den Hartog et al. 2006; Kuiper et al. 2006). 

In particular, fitting a log-parabolic functions to the hard X-ray
spectrum of \uu\ (Kuiper et al. 2006; Rea et al. 2007b; den Hartog et al. 2008) as an example resulted in a peak energy of  $279^{+65}_{-41}$ keV (den Hartog et al.  2008). This kind of
spectral model has been successfully applied to many pulsars
such as the Crab or Vela (Kuiper et al. 2001; Massaro et al. 2006a,b; Rea et al. 2007b).
Such log-parabolic spectra can be approximately obtained when
relativistic electrons are accelerated by some mechanism and
competitively cool via synchrotron or by inverse Compton scattering
losses. The narrow energy range of each log-parabolic component might then
reflect a tight balance between cooling and acceleration in a relatively
confined emission locale.

%
%
On the other hand, in some cases the hard X-ray tail at $>10$ keV can be equally well-modeled with a flat power-law with
an exponential cutoff (e.g., den Hartog et al. 2007, 2008) as opposed to
a log-parabolic form.  
One possibility suggested by this is that resonant inverse
Compton scattering by a population of highly relativistic electrons energized at altitudes
below around ten stellar radii
may provide this hard X-ray component of magnetars (see Thompson
\& Beloborodov 2005; Baring \& Harding 2007; Nobili, Turolla \& Zane
2008), 
%
%
probably using seed thermal photons emanating from the stellar
surface. In this scenario, the \fermi-LAT and {\em COMPTEL} spectroscopic constraints,
%
%
implying a turnover around $200 - 500$keV,
profoundly limit a combination of the Lorentz factor of the radiating
electrons and the typical viewing angle of the observer 
(Baring \& Harding 2007).  
Accordingly, phase-resolved spectroscopy 
will provide important diagnostics on more refined models of such a scenario (see e.g., den Hartog et al. 2008).
Note that Tr\"umper et al. (2010) recently invoked a bulk-Comptonization,
fallback disk model as an alternative, non-magnetar explanation for these tails.

The low \fermi-LAT upper bounds provide interesting constraints on
postulated magnetar synchrotron/curvature emission from high altitudes. The
emerging paradigm for young pulsars that are 
%
%
bright in the 100 MeV -- 10 GeV energy range (Abdo et al. 2010c) is that they 
emit due to 
acceleration in a slot-gap or outer-gap potential not far
from their light cylinders. Much earlier, Cheng 
\& Zhang (2001) and Zhang \& Cheng (2002) proposed an outer-gap model for
magnetar emission above 30 MeV, mediated by pairs created at high altitudes
in collisions between X-rays originating on or near the surface, and
GeV-band primary photons from electrons accelerated in the gap.  
Given the nominal \fermi-LAT sensitivity, their model predicted that 
\sgrb\, and five AXPs (see Fig.~5 of Cheng \& Zhang 2001)
would have been observable within a year with fluxes  of
the order of $10^{-7}-10^{-9}$ photons~cm$^{-2}$~s$^{-1}$, depending
on the assumed parameters. However, \fermi-LAT does not detect any
of these magnetars in 17 months of data.
This strong observational diagnostic necessarily forces a revision of the
parameter space applicable for the viability of their outer gap model to each magnetar.
 
\acknowledgments

The \fermi-LAT Collaboration acknowledges support from a number of agencies and institutes for both development and the operation of the LAT as well as scientific data analysis. These include NASA and DOE in the United States, CEA/Irfu and IN2P3/CNRS in France, ASI and INFN in Italy, MEXT, KEK, and JAXA in Japan, and the K.~A.~Wallenberg Foundation, the Swedish Research Council and the National Space Board in Sweden. Additional support from INAF in Italy and CNES in France for science analysis during the operations phase is also gratefully acknowledged.

\end{document}